\documentclass[%
 reprint,
superscriptaddress,
 amsmath,amssymb,
 aps,
floatfix,
]{revtex4-2}

\usepackage{graphicx}
\usepackage{dcolumn}
\usepackage{bm}
\usepackage{xcolor}
\usepackage[normalem]{ulem}

\usepackage[
margin=0.9in,
]{geometry}

\usepackage[utf8]{inputenc} 
\usepackage[parfill]{parskip} 

\begin{document}

\title{Towards a Scalable Linear-Cavity Enhanced Warm-Vapour Photonic Quantum Memory}
\author{Bharath Srivathsan}
  \email{bharath@orcacomputing.com}
  \affiliation{ORCA Computing Ltd., LG, 30 Eastbourne Terrace, London W2 6LA, United Kingdom}
\author{Rafal Gartman}
  \affiliation{ORCA Computing Ltd., LG, 30 Eastbourne Terrace, London W2 6LA, United Kingdom}
\author{Robert J. A. Francis-Jones}
  \affiliation{ORCA Computing Ltd., LG, 30 Eastbourne Terrace, London W2 6LA, United Kingdom}
\author{Peter J. Mosley}
  \affiliation{ORCA Computing Ltd., LG, 30 Eastbourne Terrace, London W2 6LA, United Kingdom}
  \affiliation{Centre for Photonics and Photonic Materials, Department of Physics, University of Bath, Bath, BA2 7AY, United Kingdom}
\author{Joshua Nunn}
  \affiliation{ORCA Computing Ltd., LG, 30 Eastbourne Terrace, London W2 6LA, United Kingdom}
\date{\today}

\begin{abstract}
The coherent storage, buffering and retrieval of photons in a quantum memory enables the scalable creation of photonic entangled states via linear optics and repeat-until-success, unlocking applications in quantum communications and photonic quantum computing. Quantum memories based on off-resonant cascaded absorption (ORCA) in atomic vapours allow this storage to be broadband, noise-free, and high efficiency. Here, we implement a \emph{cavity-enhanced} ORCA memory with reduced footprint and reduced power requirements compared to conventional single-pass schemes. By combining a strong magnetic field with polarisation control, we maintain a Doppler-free interaction and eliminate the need for optical pumping. Our design establishes the feasibility of large arrays of ultra-compact, low-power, near-unit-efficiency, noiseless quantum memories running at GHz bandwidth, without the need for atom trapping or cryogenics.
\end{abstract}

\maketitle

\break

Scalable quantum memories with high efficiency and bandwidth are a critical requirement for realising a large-scale photonic quantum computer~\cite{Heshami12112016, Slussarenko2019, SHINBROUGH2023297,Lei:23}. Alkali atoms in a warm vapour cell have proved to be one of the most promising quantum memory platforms due to room-temperature operation, high efficiency and signal-to-noise and bandwidth offering, and no active vacuum infrastructure requirements~\cite{Guo2019, PhysRevA.100.033801,Davidson2023, hosseini:2011}. A standard atomic-ensemble-based quantum memory is implemented in a {\em single-pass} configuration in a vapour cell that is a few centimetres long \cite{Reim:2010aa}. A control field produced by a high-power laser is needed for operating the memory, and additional lasers for state preparation of the atoms. The number of required laser systems, optical power levels, complex beam orientations, and large physical dimensions of the sub-components makes these single-pass devices hard to scale up to the numbers needed in a fault-tolerant quantum computer. 

The interaction strength between light and atoms can be drastically enhanced by using an optical cavity to bring down the power requirements and physical size of the devices. A standard linear optical cavity formed by two retro-reflecting mirrors supports a standing wave optical mode. However, a standing wave mode is not suitable for a room temperature atomic ensemble based quantum memory since the atoms are in constant motion and experience a time varying field as they move between the nodes and anti-nodes of the standing wave, significantly reducing memory lifetime. Moreover, resonant enhancement for both signal and control fields in this type of cavity results in photon storage via two-photon transition for both co- propagating and counter- propagating beam orientations of the two fields. This further reduces memory efficiency as most memory protocols work efficiently only for one of these orientations due to effective Doppler cancellation. Instead of the linear design, travelling-wave cavities with ring or bow-tie geometries have been used to avoid these problems~\cite{Ma2022,PhysRevLett.116.090501}. However such non-trivial geometries are difficult to scale. Additionally, high acceptance bandwidth for the signal -- on the order of 1\,GHz -- even for moderate cavity finesse requires sub-millimetre mirror spacing which is challenging for a ring cavity. A linear cavity, on the other hand can be engineered for sub-mm spacing with mirrors on optical fiber ends ~\cite{Steinmetz2006StableFF,Saavedra2021,Gulati2017} or micro-fabricated mirrors~\cite{Jin2022}. Miniature vapour cell technology via micro-fabrication has matured significantly in the past decades~\cite{Wang2022}, and quantum memory has recently been demonstrated in a micro-fabricated vapour cell~\cite{PhysRevLett.131.260801}. A quantum memory implemented within a miniature linear cavity could be scaled to large arrays needed for quantum computing and networking applications, but has not been realised thus far.

\begin{figure*}[t!]
\begin{center}
\includegraphics[width=0.95\textwidth]{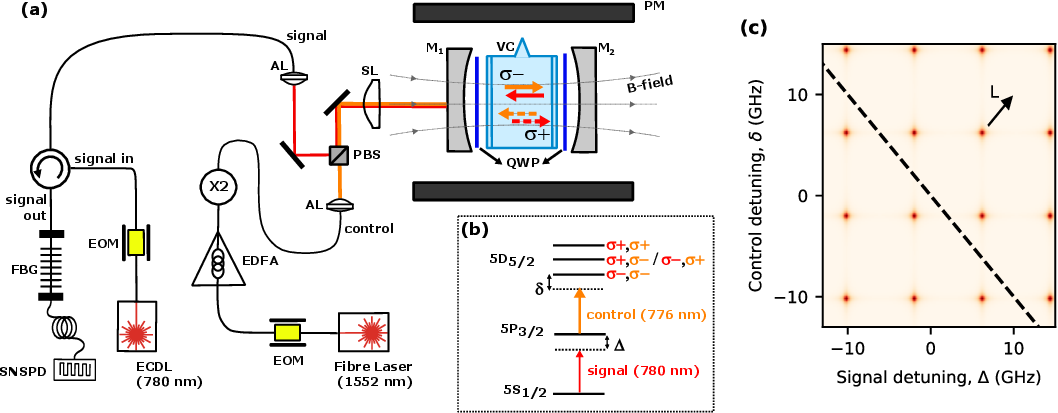}
\caption{(a) Schematic of the experimental setup indicating the key components. AL: Aspheric fibre collimation lens of focal length 4.5mm; SL: Plano-convex singlet lens of focal length 100\,mm used for mode-matching the signal and control beams to the TEM$_{00}$ mode of the cavity; PBS: Plate polarization beam splitter used to combine the signal and the control beams; M$_{1}$: Cavity in-coupling mirror with anti-reflection coating on the planar side and high reflection coating on the concave side with a reflectivity of 60\%; M$_{2}$: Cavity end mirror with a reflectivity 99.98\% for the concave side; QWP: Quarter wave plates used to rotate the polarization by 90 degrees upon each reflection in the cavity; VC: Vapour cell containing isotopically pure $^{87}$Rb; PM: Ring shaped permanent magnets used to produce axial magnetic field along the propagation direction $z$.  The input signal and control couples in to the cavity via the mirror M$_{1}$. The signal output couples through the same optical fibre. An optical circulator is used to separate the output signal from the input and sent to the SNSPD for detection.
Polarization of the intra-cavity fields are circular and correlated to their propagation directions. Field Polarization driving $\sigma-$ transitions in the atoms are indicated using solid lines and that driving $\sigma+$ transitions are indicated with dashed lines. (b) Simplified level scheme. Transition frequencies for dipole allowed transitions with polarisation combination other than ($\sigma-$, $\sigma-$) for signal and control fields are shifted off-resonance by the strong B-field as indicated in the level scheme. (c) The 2D plot shows modelling of the signal-control resonances for the design parameters used in our experiment; The signal and control frequencies are expressed in terms of detunings $\Delta$ and $\delta$ from the respective atomic transitions. The darker (red-orange) circular spots indicate the signal-control frequency pair for which the dual resonance criteria is satisfied. The black dashed line indicates the parametric line that satisfies the two-photon resonance \emph{i.e.,} $\delta$\,=\,$-\Delta$. The arrow labelled L indicates the direction in which the resonances shift when the mirror spacing is changed by temperature tuning. We measure a resonance shift of $\approx$\,3.2\,GHz/$^\circ$C.}
\label{fig:concept}
\end{center}
\end{figure*}

In this paper we present a ladder-type (ORCA) quantum memory~\cite{PhysRevA.97.042316} implemented in a room-temperature rubidium vapour cell placed in a linear cavity. The operating concept of our experiment is illustrated in Figure~\ref{fig:concept}. A near-resonant single photon signal field and a strong control field causes a two-photon absorption in the atomic vapour, thereby storing the photon as a spin-wave excitation. The cavity is tuned such that both the signal field and the control fields are resonant with the fundamental transverse (TEM$_{00}$) mode of the cavity, and frequencies of these two fields sum to the atomic two-photon transition frequency. To avoid standing wave resonance of the fields in the cavity, we use intra-cavity quarter-wave plates (QWP) before each retro-reflection. A circularly polarised field in the forward propagating ($+z$) direction becomes orthogonally polarised while propagating in the backward ($-z$) direction, thus preventing intensity standing waves. Another issue with the linear cavities is the presence of both co-propagating and counter-propagating signal-control beam orientations as depicted in Figure~\ref{fig:concept}(a). To cause a selective storage of only one of these orientations, we apply a strong magnetic field to increase the frequency separation between the magnetic-spin sub-levels, and use polarisation selection rules to allow absorption of only the desired beam orientation \emph{i.e}, counter-propagating for ORCA memory. The applied field strength is strong enough to separate the two-photon transition frequencies for orthogonal spin transitions to be higher than the excitation bandwidth. If the frequency and polarisation of the signal and control fields are tuned to excite the two-photon resonance for one of the beam orientations, the transitions for other beam orientations will be off-resonance, suppressing this loss mechanism. The method used for tuning the cavity to satisfy double resonance, that is, cavity resonance for both the signal and control fields, while also matching the two-photon transition frequency, is illustrated in Figure~\ref{fig:concept}(c). We use temperature to tune the length of the cavity to hit the two-photon resonance, and choose an appropriate longitudinal cavity mode that provides sufficiently large intermediate state detuning ($\Delta$) compared to the Doppler-broadened linewidth of $\Gamma\,=\,2\pi\times$0.55\,GHz to reduce signal loss due to linear absorption. In the experiment, we work with a cavity mode that provides $\Delta$ of $-2\pi\times8\,$GHz.

\begin{figure*}[t!]
\begin{center}
\includegraphics[width=16cm]{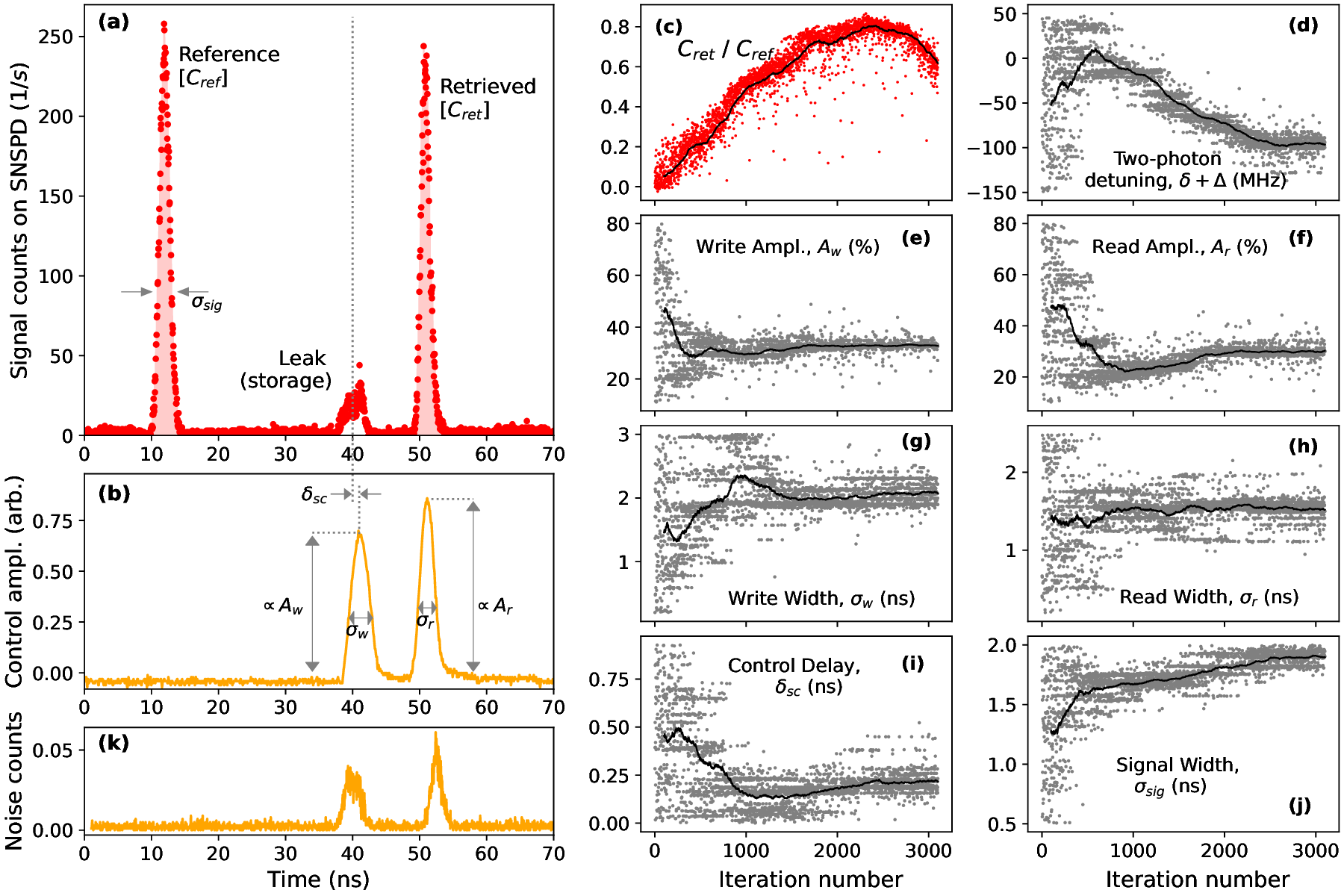}
\caption{The detection of signal pulses on the SNSPD is shown in plot (a), and the corresponding control pulse detection using a fast photodiode and an oscilloscope is shown in plot (b). In the signal mode, a {\em reference} pulse followed by an identical storage pulse is sent to the memory. A {\em write} control pulse overlapping in time with the storage pulse results in absorption of the signal into the memory. The unstored signal is detected as {\em leak pulse} by the SNSPD. After a storage time of 12.5\,ns a {\em read} control pulse is sent to the memory which retrieves the stored signal and is detected as {\em retrieved pulse} by the SNSPD. The ratio of total counts in the retrieved pulse to the reference pulse ($C_{ret}/C_{ref}$) is proportional to the memory efficiency, and is used as a objective function for maximising the efficiency. The results of a real-time optimisation run is shown in the plots (c) to (j). It can be seen that value of the objective function in (c) initially grows with iteration number as the various experimental parameters steer towards their optimal values, but the cavity resonance drift prevents convergence. The features in the measurement corresponding to some of these parameters are annotated in the plots (a) and (b). The pulse widths in (g), (h) and (j) are defined at full-width-half-maximum. The two-photon detuning in (d) is tuned using control detuning $\delta$. At the start of the optimisation, we tune the control laser frequency to hit the two-photon resonance \emph{i.e.,} $\delta$\,=\,$-\Delta$. (k) Noise counts in the signal mode for the same integration time used in (a). The plot shown is obtained from average of counts measured for 1000 traces.}
\label{fig:optimisation}
\end{center}
\end{figure*}

The signal in our experiment is a weak coherent pulse generated from a 780\,nm external cavity diode laser (ECDL), and pulse-carved into a Gaussian shaped temporal pulse using an Electro-Optic amplitude Modulator (EOM). The control field is generated from a 1552\,nm continuous-wave fibre laser, pulse-carved using an EOM, amplified using an Erbium Doped Fibre Amplifier (EDFA), and frequency doubled to 776\,nm.  The optical cavity is formed of two plano-concave mirrors $M_1$ and $M_2$ with radius of curvature of 15\,mm and reflectivity of 0.6 and 0.9998. The reflectivities were chosen to achieve a cavity linewidth of approximately 1\,GHz with practically achievable mirror spacing of 18\,mm. The measured free-spectral-range $\nu_c$ of our cavity is $8.3\pm0.2$\,GHz, and linewidth $\kappa$ is 0.88$\pm$0.06\,GHz, resulting in the cavity finesse $\mathcal{F}$ of 9.5$\pm$0.7 (see Appendix). The signal and control fields from polarisation maintaining (PM) single-mode optical fibres are coupled to the cavity via the first cavity mirror \emph{i.e}, the one with lower reflectivity. Circular polarized intra-cavity field is used for both signal and control fields, and an axial magnetic field of 169$\pm$3\,mT applied using a ring shaped Neodymium permanent magnet defines the system quantization axis. An isotopically pure $^{87}$Rb vapour cell (VC) with internal optical path-length ($L$) of 6\,mm is placed between the cavity mirrors. The cavity - vapor cell assembly is temperature stabilized to approximately 85\,$^\circ$C using a resistive heater resulting in a single pass \emph{static} optical depth $d\approx 200$ for the signal field. The optical depth and the cavity finesse puts the cooperativity parameter of our system $C=2\,d\,\mathcal{F}$ at about 3800~\cite{PhysRevA.76.033804}, with the calculated internal efficiency approaching unity.

The experimental sequence begins with a signal pulse with a mean photon number of $\approx\,0.8$, and a temporally overlapped write control pulse sent to the cavity. The signal pulse is partially absorbed by the atoms resulting in storage in the memory. The unabsorbed signal leaks out of the cavity and coupled back into the same PM fibre that delivers the input signal pulse. An optical circulator is used to spatially separate the leaked signal from the input signal and sent to superconducting nanowire single photon detector (SNSPD) for detection. After a desired storage time, a read control pulse is sent to the cavity memory that retrieves the stored signal from the memory. The retrieved signal is emitted in the same mode as the leaked signal and is detected by the same SNSPD at a delayed time corresponding to the storage time in the memory. Histogram of signal detection events on the SNSPD recorded using a Time-to-Digital Convertor is shown in Figure~\ref{fig:optimisation}. A reference signal pulse with the control beams turned off is used to determine the input photon counts and used for normalization.

A number of experimental parameters influence the efficiency of the cavity memory system, including the pulse energy and pulse width of the write, read, and signal fields, as well as the temporal and spectral overlap of the signal and control pulses. In our experiment all these parameters are configured to be controlled remotely using a computer. To find the optimal values of these pulse parameters for the highest retreival efficiency, we run a real time optimization routine. Ratio of the integrated SNSPD counts in the retrieved signal pulse to the reference signal pulse is used as a objective function. The optimisation runs in a feedback loop in which experimental parameters are varied based on the feedback from the measurement of this objective function. We tested various algorithms for searching for optimal parameters and found that applying non-dominated sorting genetic algorithm (pymoo.algorithms.moo.nsga2) resulted in the best performance~\cite{pymoo, 996017}. A similar technique has recently been reported for optimising a $\Lambda$-type quantum memory~\cite{PhysRevApplied.22.024026}. Result from an optimization run is shown in the Figure~\ref{fig:optimisation}. 
The algorithm improves the objective function by steering the parameters towards their optimal values. However, we find that the cavity resonance drifts due to temperature instability which prevents convergence~\cite{suppl}. Since this resonance frequency is currently not a tunable parameter in our optimization, the objective function starts to drop at higher iteration numbers. 
The memory efficiency is determined from the objective function by including the intra-cavity losses $\zeta$ for the reference counts: $\frac{C_{ret}}{C_{ref}/(1-\zeta)}$. This loss is independently measured to be 0.68$\pm$0.04 (see Appendix), resulting in best measured memory efficiency of 27$\pm$3\,\% for a storage time of 12.5\,ns. The error in efficiency is dominated by the error in measuring $\zeta$. The intrinsic efficiency \emph{i.e.,} efficiency if the intra-cavity loss $\zeta$ can be eliminated, is predicted to be $\approx$\,90\% by numerical methods (see supplementary material~\cite{suppl}). This is significantly lower than the near-unit efficiency expected for our system cooperativity $C$ of 3800. This is because the optimization yields a sub-optimal write control pulse shape due to lack of phase control and pulse shaping bandwidth limitations.
Noise present in the signal mode is measured by switched off the signal and measuring the average SNSPD counts for the same integration time. A Signal-to-Noise Ratio (SNR) of 37$\pm$3\,dB is determined from the measurement shown in figure~\ref{fig:optimisation}(k), which corresponds to $3\times10^{-4}$ noise photons per pulse.

\begin{figure*}
\begin{center}
\centering
\includegraphics[width=16cm]{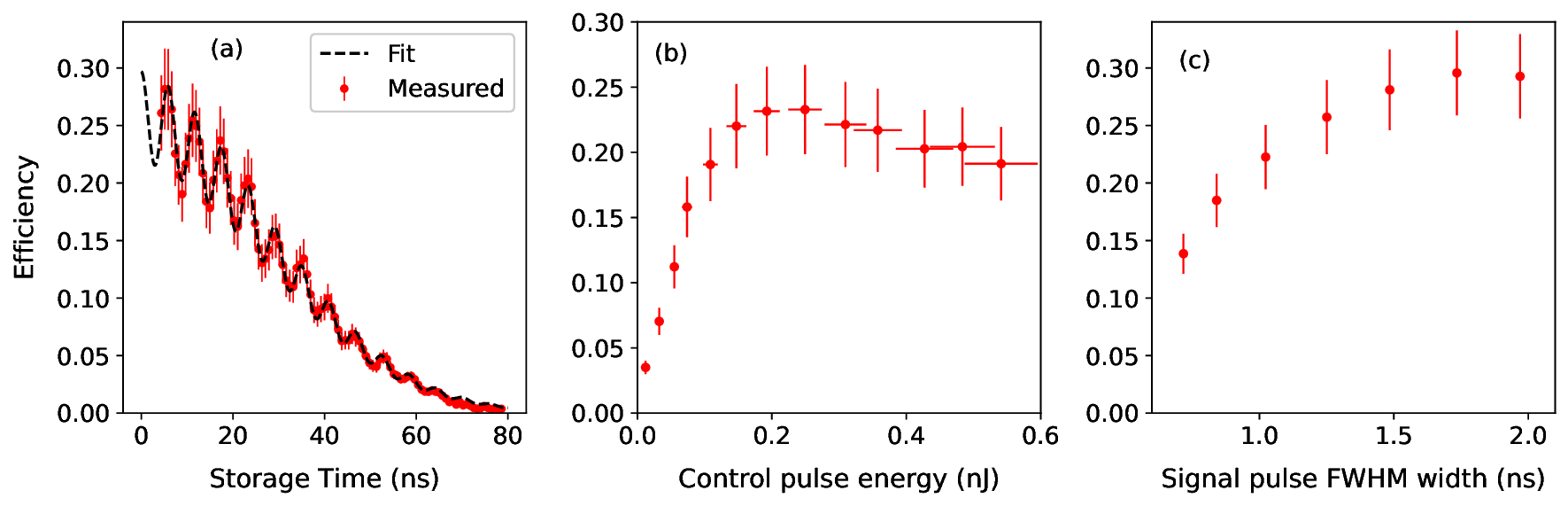}
\caption{(a) Memory efficiency measured as a function of signal storage time. The black (dashed) line shows the fit to the model described in the text. (b) Memory efficiency at 12.5\,ns storage time as a function of control pulse energy of the write pulse. The fraction of the read pulse energy to write pulse energy and the other parameters are fixed at the optimum value. The pulse energy is determined from the detection of control pulse amplitude using a fast photodetector. The error in pulse energy comes from the uncertainty in photodetector sensitivity and losses. (c) Memory efficiency vs signal pulse temporal width expressed in Gaussian Full-Width-at-Half-Maximum (FWHM).}
\label{fig:param_scans}
\end{center}
\end{figure*}

Some of the other important metrics of a quantum memory are its lifetime, acceptance bandwidth, and control power requirement. Characterization of these parameters is shown in Figure~\ref{fig:param_scans}. To determine the lifetime we measure the efficiency while varying the storage time of the signal in the memory. Since there are two addressable two-photon absorption lines within the cavity linewidth (see Appendix), the efficiency oscillates with the difference frequency. To extract the lifetime, we use a damped oscillatory fit function assuming a Gaussian model for transition broadening due to spin-wave dephasing.
\begin{equation}
\eta\,=e^{-\gamma_m\,t}\,e^{\frac{-\pi^2\,\nu'^2\,t^2}{4\,ln(2)}}|A\,+\,B\,e^{i\,\omega\,t}|^2
\end{equation}
where the natural linewidth of the $\textrm{5D}_\textrm{5/2}$ state is $\gamma_m$ is a fixed parameter with value $2\pi\times$\,0.66\,MHz~\cite{PhysRevA.78.062506}. The free parameters extracted from the fit are: the transition broadening width $\nu'$\, of 12.6$\pm$0.5\,MHz, oscillation frequency  $\omega$ of 2$\pi\times($171$\pm$2)\,MHz and relative transition strengths $A$ of 0.51$\pm$0.04 and  $B$ of 0.038$\pm$0.003. This results in a zero storage time efficiency of 30$\pm$3\,\% and $1/e$ lifetime of 39$\pm$1\,ns. The oscillation frequency and the transition broadening width from this fit is consistent with independent cw spectroscopy measurement (see Appendix). The measured lifetime is shorter than the residual Doppler limited lifetime of approximately 120\,ns due to the non-uniformity of the applied B-field. The oscillation can be readily minimised by using stronger B-field; For instance, a field magnitude of 250\,mT is expected to reduce the amplitude of these oscillations by about a factor of 10. In contrast, a standard single-pass ORCA memory requires additional high powered lasers for optical pumping to eliminate such oscillations.
The measurement of efficiency vs control pulse energy for a storage time of 12.5\,ns is shown in Figure~\ref{fig:param_scans}(b). The highest efficiency is observed for a write pulse energy of 0.2\,nJ. This pulse energy is about an order of magnitude smaller compared to an optimized single-pass ORCA memory~\cite{Finkelstein2018}. 
The memory acceptance bandwidth is determined by measuring the optimised efficiency for different signal pulse durations shown in Figure~\ref{fig:param_scans}(c). The efficiency saturates for pulse widths greater than 1.5\,ns; this requires signal bandwidths of under 0.6\,GHz to achieve efficiency within $<$5\% of the maximium. The bandwidth limitation stems from the cavity linewidth. Higher bandwidth can be achieved by using shorter cavity lengths which lowers the memory efficiency due to reduction in single pass optical depth. It is in-principle possible to prevent a reduction in optical depth by operating the vapour cell at higher temperature. Vapour cell operating temperature of 120\,$^\circ$C without degradation of optical surfaces have been reported~\cite{Zektzer:24}. We therefore estimate that an order of magnitude higher acceptance bandwidth without reduction in efficiency would be achievable.

In conclusion, we have demonstrated a warm alkali vapour quantum memory with both signal and control field interactions enhanced by a linear cavity. The cavity allows reduction in the interaction length and the control power requirement by a an order of magnitude even for a low finesse of $\approx10$. Further reduction is possible with higher finesse and a smaller mode-volume cavity if the intra-cavity losses can be reduced. The losses in our experiment arise mainly from the the vapour cell windows, and can be eliminated by placing the cavity inside the vapour cell. 
The linear design of our memory with minimal laser infrastructure requirements is well suited for integration with other established cavity and vapour cell fabrication methods~\cite{Steinmetz2006StableFF}-\cite{PhysRevLett.131.260801}, and therefore can be scaled into an array of quantum memories needed for useful, large-scale quantum technology applications. Finally, as an outlook, we note that in the ORCA interaction, a control optical field modifies the absorption or dispersion of a signal field. With a sufficiently high quality cavity, and a sufficiently small mode volume, this becomes a two-photon (cross-Kerr) interaction which could be used to deterministically entangle the signal and control at the level of individual photonic qubits~\cite{PhysRevLett.108.030502}. This opportunity motivates further work on cavity-enhanced ORCA interactions.

The data that support the findings of this article are openly available~\cite{dataset}.
\begin{acknowledgments}
We would like to acknowledge funding from innovate-UK (award number: 10102696). This work was also partially funded by the US Air Force Research Laboratory (award number : FA8655-21-1-7059).
\end{acknowledgments}

\nocite{*}
\bibliography{citations}

\clearpage
\appendix*

\section{End Matter}

{\bf Signal pulse photon number - }
We use a temporally Gaussian shaped, weak coherent state pulses as a signal instead of true single photon states. This is valid since at low signal intensity well below saturation, the quantum memory has a linear response to the input photon number in the signal pulse. We measure the transmission loss from the signal input fibre to the SNSPD to be $\approx\,13$\%. From the integrated SNSPD counts in the reference signal pulse and this transmission, we determine the mean photon number in the signal pulse at the memory input to be $\approx\,0.8$. 

\begin{figure*}[t!]
\begin{center}
\includegraphics[width=0.84\textwidth]{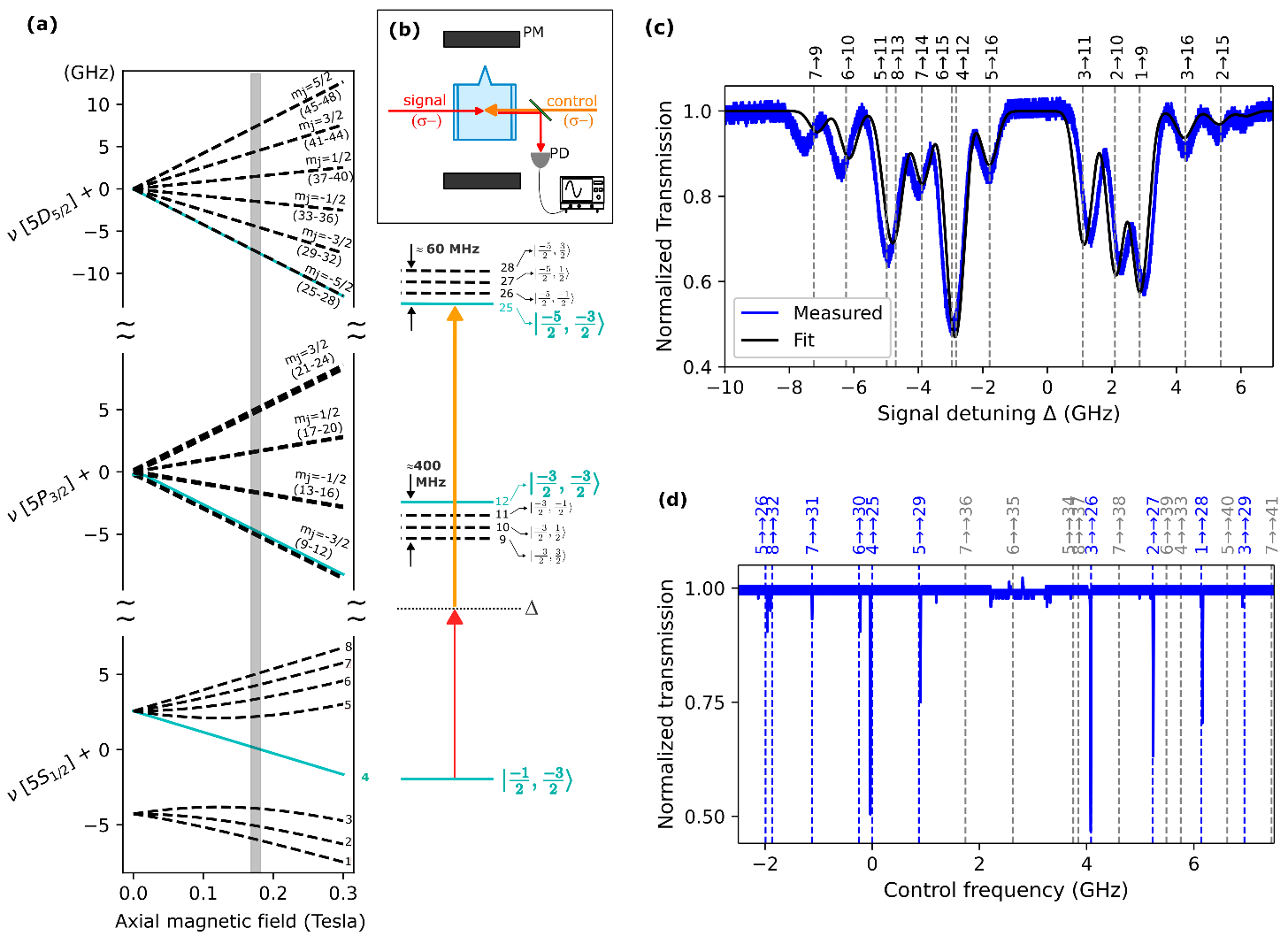}
\caption{(a) Breit-Rabi diagram with the states labelled from 1 to 48. The grey shaded region indicates the level spacings for the magnetic field used in the experiment. The relevant transitions at this magnetic field is shown on the right with the three-levels used for the memory is highlighted in cyan. The state vectors indicate the total spin and nuclear spin quantum numbers $m_j$ and $m_i$ of the states along the direction of the B-field. (b) Experimental setup used to perform spectroscopy. A weak 780\,nm cw signal beam used as a probe passes through the vapour cell, and its transmission is detected using a photodetector (PD) and an oscilloscope. A strong cw control beam at 776\,nm is used for two-photon spectroscopy. It is aligned to overlap with the signal in the vapour cell. A permanent magnet PM is used to produce the B-field. Both the control and the signal fields are produced by Toptica DLPro external cavity diode lasers (ECDL).
(c) One-photon absorption spectroscopy performed using $\sigma-$ signal beam. The control beam is blocked for this measurement. The frequency of the signal is expressed in terms of detuning from the $5S_{1/2} \rightarrow 5P_{3/2}$ central frequency shown in (a). The measured absorption dip is shown as coloured (blue) line. Model fit to theory indicated in black solid line is used to determine the B-field strength to be $169\pm3$\,mTesla. The only other fit parameters include frequency offset, and atomic vapour optical depth.  The Doppler broadening width and the relative transition strengths of the lines are fixed according to the theory. The deviation of the measured dip from the theory is mainly due to non-linear response of the piezo voltage scan used for frequency tuning the laser, which leads to systematic error in the attributed frequency of the probe beam.
(d) Two photon absorption spectrum of the signal beam measured by scanning the frequency of the control laser. The frequency of the signal is fixed to a frequency corresponding to 0\,GHz in (c). The polarizations of both signal and control were set to $\sigma-$ for this measurement. The vertical dotted lines indicate strong absorption lines from theory. The blue lines indicate transitions with $\sigma-$ polarization for both signal and control fields which is the configuration used for memory. The grey lines indicate the transitions for other pairs of polarization that would result in signal loss in the cavity. Refer to supplementary material for the relative transition strengths and detunings~\cite{suppl}. 
}
\label{fig:spectroscopy}
\end{center}
\end{figure*}

{\bf One and two photon spectroscopy - }
The operating principle of our cavity enhanced quantum memory relies on energy level shifting via strong magnetic fields, and selective two-photon absorption in the cavity for the desired beam orientation implemented by polarization selection rules. Some of the strong two-photon absorption lines that would be suitable for the desired beam orientation also have dipole allowed transitions closely spaced in energy for the other beam orientations resulting in signal loss. It is therefore important to identify the suitable lines in that would work for an efficient memory operation while not having these loss channels.  The theoretical energy shifts are obtained by exact diagonalization of the Zeeman interaction hamiltonian and the corresponding relative transition strengths calculated from the Clebsch-Gordan coefficients. The calculated energy shifts of the relevant levels as a function of magnetic field is shown in Figure~\ref{fig:spectroscopy}(a). The sub-levels are indexed from 1 to 48 for identification and referencing in spectroscopic measurements. The transition that we choose to implement the memory is highlighted in colour. We choose this transition since it is a strong transition with no loss channels within a few GHz.
The experimental setup used and the results of the spectroscopy measurements are shown in Figure~\ref{fig:spectroscopy}(b). The first step was to measure the B-field strength from the permanent magnet at the position of the atoms. We perform Doppler-broadened absorption spectroscopy on the D2 line using a weak 780\,nm probe beam shown in shown in Figure~\ref{fig:spectroscopy}(c). The measured absorption dip is fit to the theory to extract the field magnitude. Next, we perform two-photon spectroscopy on the $5S_{1/2} - 5P_{3/2} - 5D_{5/2}$ transition using a strong 776\,nm control beam and a weak 780\,nm probe. The result of this measurement is shown in Figure~\ref{fig:spectroscopy}(d). All the strong dipole allowed transitions and their frequencies from the theory are overlaid as dotted vertical lines. The blue lines indicate the transitions for the desired beam orientation, while the gray ones are the loss channels i.e., transitions corresponding to the other beam orientations. 

\begin{figure}[t]
\includegraphics[width=0.84\columnwidth]{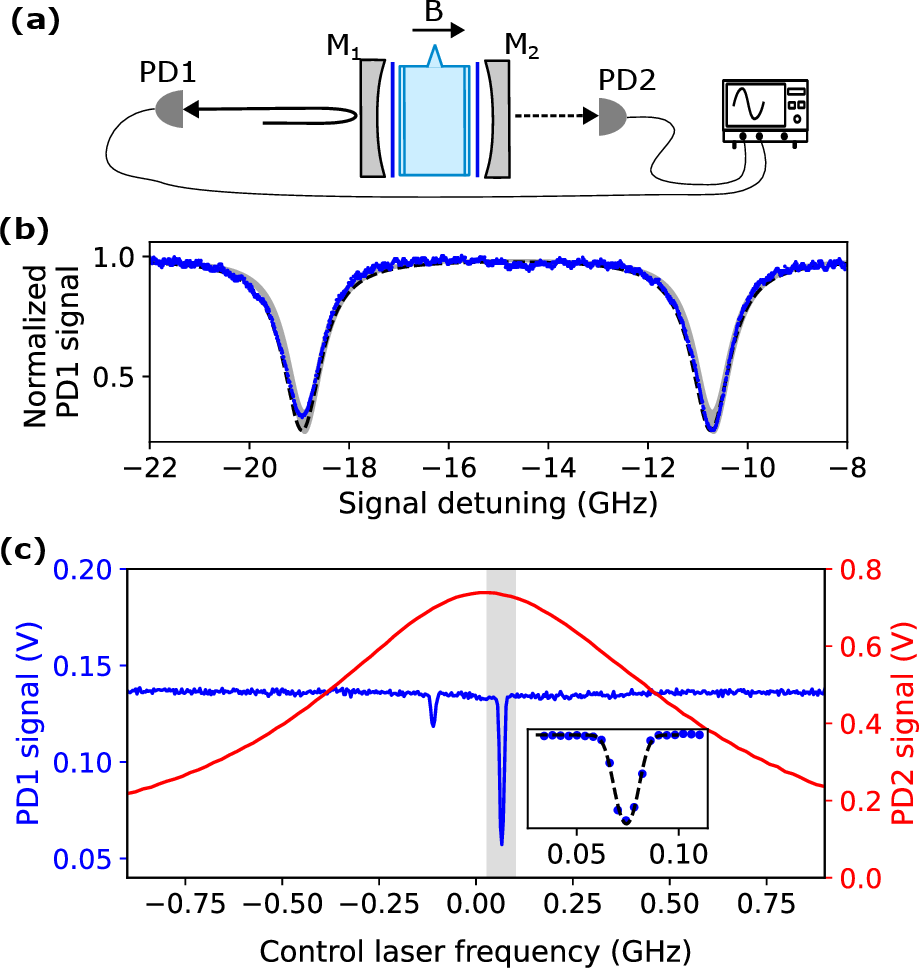}
\caption{(a) Setup used to characterise the cavity performance. The cavity response is measured using a signal and/or control laser in cw operation in reflection and transmission using photodetectors PD1 and PD2 respectively. (b) Frequency response of the cavity in reflection measured using a weak cw signal laser shown in blue. The 776\,nm control laser is switched off for this measurement. The frequency is expressed in terms of detuning ($\Delta$) from $5S_{1/2} \rightarrow 5P_{3/2}$ central frequency shown in Figure~\ref{fig:spectroscopy}(b). Curve fit to the model described in the text is shown as black dashed line with the shaded grey areas showing the uncertainty due to parameter errors. 
(c) Frequency response with a weak signal laser and a strong control laser. Signal frequency is tuned to be on-resonance with the cavity and an atomic intermediate state detuning of $\approx\,-11$\,GHz in (b). The control frequency is scanned and its transmission through the cavity measured on PD2 shown as red line in the plot. The signal detection on PD1 shown as blue line reveals two narrow two-photon absorption lines occurring within the cavity bandwidth with a frequency separation of $\approx$175\,MHz.(d) Zoomed in to show the shaded region in (c) i.e., the stronger of the two absorption lines. The black dashed line is a fit to Gaussian broadened absorption line used to determine the FWHM width of 11.8$\pm$2.4\,MHz.
}
\label{fig:cavity_characterisation}
\end{figure}

{\bf Cavity characterisation - }
Cavity Free-Spectral-Range (FSR), linewidth, and excess optical loss ($\zeta$) are some of the important metrics necessary to quantify the performance of our cavity memory.  We send a cw laser at 780\,nm via the same optical path used for the signal pulses to perform this measurement. We scan the frequency of the laser and measure the transmitted and reflected powers using a photodiode. The results of the frequency scan measurements are shown in figure~\ref{fig:cavity_characterisation}
Using the following model function for reflection response from the cavity, we perform a curve fit to the measured reflection data with Free-Spectral-Range $\mathcal{F}$, and {\em round-trip} excess optical loss $\zeta_{rt}$ as free parameters. The reflectivities $R_1$ and $R_2$ are fixed to the value from the manufacturer test report of 0.6 and 0.9998 respectively.
\begin{equation}
R(\Delta)\,=\,|\sqrt{R_1} - \frac{(1-R_1)\,\sqrt{R_2\,(1-\zeta_{rt})}\,e^{i\,2\pi\Delta/\nu_c}}{1-\sqrt{R_1\,R_2\,(1-\zeta_{rt})}\,e^{i\,2\pi\Delta/\nu_c}}|^2
\end{equation}
The free parameter values derived from the fit are: $\nu_c$ = $8.3\pm0.2$\,GHz, and $\zeta_{rt}$ = $13.5\pm1.5\,\%$. This corresponds to an effective cavity finesse of $\mathcal{F}$\,=\,$9.5\pm0.7$, cavity linewidth of $\kappa$\,=\,$0.88\pm0.06$\,GHz and on-resonance excess optical loss $\zeta$ for a cw signal of $-5\pm0.6$\,dB. Note that $\zeta$ is the total cavity insertion loss which is different from the round-trip excess loss $\zeta_{rt}$. This excess loss comes from scattering off the different optical surfaces within the cavity. The loss from the waveplates alone is independently determined to be negligible; Hence, we deduce that the excess loss is dominated by scattering losses at the vapour cell windows due to imperfect anti-reflection coatings and rubidium condensing on the windows.

\end{document}